\documentclass[preprint,showpacs,preprintnumbers,amsmath,amssymb]{revtex4}

\usepackage{graphicx}
\usepackage{natbib}
\usepackage{multirow}
\usepackage{latexsym,amssymb}
\usepackage{amsmath}

\newcommand{\Rpp}{\vec{R}_{\perp}}
\newcommand{\kpp}{\vec{k}_{\perp}}
\newcommand{\kpr}{\vec{k}_{\parallel}}
\newcommand{\rpr}{\vec{r}_{\parallel}}

\newcommand{\ver}{\vec{r}}
\newcommand{\veR}{\vec{R}}
\newcommand{\vek}{\vec{k}}

\begin{document}
\bibliographystyle{plainnat}

\title{Influence of rotational instability on the polarization structure of SrTiO$_3$}
\author{Yanpeng Yao}
\author{Huaxiang Fu}
\affiliation{Department of Physics, University of Arkansas,
Fayetteville, AR 72701, USA}
\date{\today}

\begin{abstract}
The $k$-space polarization structure and its strain response in
SrTiO$_3$ with rotational instability are studied using a combination of
first-principles density functional calculations, modern theory of
polarization, and analytic Wannier-function formulation. (1) As one outcome
of this study, we rigorously prove---both numerically and
analytically---that folding effect exists in
polarization structure. (2) After eliminating the folding effect, we
find that the polarization structure for SrTiO$_3$ with rotational
instability is still considerably different from that for
non-rotational SrTiO$_3$, revealing that polarization structure is
sensitive to structure distortion of oxygen-octahedra rotation and
promises to be an effective tool for studying material properties.
(3) Furthermore, from polarization structure we determine the microscopic
Wannier-function interactions in SrTiO$_3$. These interactions are found
to vary significantly with and without oxygen-octahedra rotation.
\end{abstract}


\maketitle

\section{Introduction}

Materials with perovskite or perovskite-based structures have attracted
wide attention for their varieties of properties in
ferroelectricity, magnetism, superconductivity, etc.
In these perovskite solids, an important class of structural
distortions that deviate them from ideal (cubic) perovskites
is the rotation of the oxygen octahedra, which may considerably alter
the physical and chemical properties of materials.
For instance, rotation of oxygen octahedra
in incipient ferroelectric SrTiO$_3$ was found to couple strongly with
tetragonal strain, and consequently, affect significantly the structural
and dielectric behaviors below 105K.\citep{Uwe,Muller}
Octahedral rotation also plays a key role in improper
ferroelectrics such as YMnO$_3$, where structural instability due to
zone-boundary phonon mode couples with electric polarization\citep{Fennie},
and this coupling was recently shown to
exhibit an interesting absence of critical thickness in
improper-ferroelectric ultrathin films.\citep{Sai_1}
On magnetism, Mazin and Singh have revealed that one
main reason responsible for the drastically different magnetic
properties in two chemically similar compounds CaRuO$_3$ and
SrRuO$_3$ (where CaRuO$_3$ is paramagnetic, but SrRuO$_3$ is
ferromagnetic) consists in a larger rotation angle of RuO$_6$ octahedra in
CaRuO$_3$.\citep{Mazin}  Similarly, Fang and Terakura found that perovskite
superconductor Sr$_2$RuO$_4$ could be turned from non-magnetic to
ferromagnetic by simply increasing the rotation angle of the
RuO$_6$.\citep{Fang} Recently, Ray and Waghmare have shown that
rotation is the predominant structural instability in a
series of biferroics (LuCrO$_3$, YCrO$_3$, LaCrO$_3$, BiCrO$_3$, and
YCrO$_3$), and is closely correlated with their magnetic properties.\citep{Ray}
Based on the facts that (1) rotation of oxygen octahedra affects
various material properties, and (2) the oxygen-octahedra rotation is
common, understanding of the physics caused by oxygen rotation
is thus important.

The modern theory of polarization\citep{King-Smith,Resta} has become
important in quantitatively determining the electric and dielectric
properties of materials within first-principles density functional calculations.
According to this theory\citep{King-Smith,Resta}, the change of electric
polarization is a physically meaningful quantity and the electronic
contribution is determined by an integration of the geometric phase of
the occupied multi-fold electron Bloch wavefunctions.
Based on the modern theory of polarization,
Yao and Fu recently introduced the concept of
``polarization structure''\citep{Yao_Pol} which may provide a new scheme
to study the electronic properties of dielectric materials,
and they went on to formulate a
theory using localized Wannier functions to investigate the polarization
dispersion structure of ferroelectrics with spontaneous polarization.
More specifically, polarization structure describes the electronic phase angle $\phi$
as a function of $\kpp$ point which lies on a $\vec{k}$-plane perpendicular
to the direction of the considered polarization component.
At each $\kpp$, the phase angle is determined by the electron
wavefunctions as\citep{King-Smith,Resta}
\begin{equation}
 \phi(\kpp)=i\sum_{n=1}^M\int_0^{G_{\parallel}}d\kpr\langle
u_{n\vek}|\frac{\partial}{\partial k_\|}|u_{n\vek}\rangle \, \, ,
\label{EPhi_1}
\end{equation}
where subscripts $\parallel $ and $\perp$ indicate, respectively, the directions
parallel and perpendicular to the direction of the polarization.
$u_{n\vek}$ is the Bloch part of electron wave function,
and M is the number of occupied bands. As significant as electronic band structure,
polarization structure reveals how individual strings of different $\kpp$s contribute
to the electronic polarization, which thus offers microscopic insight
on the polarization properties. Polarization structure can be changed
by applied electric fields, in a similar spirit as electronic state in
band structure can be changed by photoexcitation.
More discussions on the importance of the polarization structure were
given in Ref.\onlinecite{Yao_Pol}.
It was also shown that the polarization structure is determined by
the interaction among Wannier functions in neighboring cells.\citep{Yao_Pol}
Polarization structure of a dielectric system may thus enable scientists to
study the microscopic interaction among Wannier functions in real
space, which is of fundamental relevance in revealing material properties.

In this work we address a topic which concerns both the rotational
instability of oxygen octahedra and polarization structure,
by investigating how the polarization dispersion structure
in ferroelectric perovskites is influenced by oxygen rotation. We will use
SrTiO$_3$ (ST) as an example. ST is known as an incipient ferroelectric,
which exhibits no polarization at the strain-free ground state but is
ready to develop a polarization under small external strains.\citep{Antons}
It was observed experimentally that
unstrained ground state of ST has a nonzero oxygen-octahedra
rotation around the (001) direction. Later, we will see from
our numerical simulations that by applying in-plane compressive strain, the rotation
angle in ST increases and the structure with oxygen rotation remains to
be the most stable ground state as the strain is small or moderate.
But at large strains, the tetragonal phase of zero rotation
becomes more stable. Comparing SrTiO$_3$ (with oxygen
rotation) with PbTiO$_3$ (of no oxygen rotation), one
anticipates both the atom-atom interaction and electron wave
functions to be altered in two materials.
Since the polarization structure is determined by
electron wavefunctions and their interaction, it is thus
of interest to study how the polarization structure changes as a
result of the octahedral rotation. It worths
pointing out that the influence of rotational instability on the
polarization structure has not been studied before\citep{Yao_Pol},
and turns out to offer interesting new knowledge on the phase
contribution at individual $\kpp$ points as well as on microscopic
interaction.

The purpose of this work is three folds: (i) To determine and reveal
how the polarization structure may look like in ferroelectrics with
rotation instability, using first-principles density functional
calculations; (ii) By comparing the polarization structure in
SrTiO$_3$ with oxygen rotation (labeled as R-SrTiO$_3$) with
the counterpart in non-rotational SrTiO$_3$
(to be denoted as NR-SrTiO$_3$ where the
oxygen-octahedral rotation is deliberately suppressed by
symmetry constraint), we intend to investigate the effect
of rotational instability on the formation of polarization
dispersion structure;
(iii) Since octahedral rotation couples with, and can
be changed by, inplane strains, we thus would like to explore
how the polarization structure in perovskites with rotation
instability may be modified by the inplane strain.

The content of this paper is organized as following. In Sec.II we
describe the computational methods used in this study,
along with a brief result on the structural responses in
R-SrTiO$_3$ and NR-SrTiO$_3$.
In Sec.III  we present the polarization structures for
R-SrTiO$_3$ and NR-SrTiO$_3$, and identify the
differences and similarities between these polarization structures.
Two main reasons contributing to the differences are revealed.
In addition, analytical formulation is also made in Sec.III to deduce from polarization dispersion
the microscopic interaction among Wannier functions.
We find that the 3rd nearest neighbor (3NN) approximation
is necessary in order to yield an accurate polarization structure
for NR-SrTiO$_3$ under large inplane strains. Based on our analytical formulation and numerical
simulation, we also describe a renormalization-group scheme
to determine the higher-order interactions between Wannier functions.
Sec.IV summarizes the results of this work.

\section {Methods and structural response}

We use first-principles density-functional theory (DFT) within local
density approximation (LDA) \citep{LDA} to determine
cell parameters and internal atomic positions in zero-strained
SrTiO$_3$ as well as in SrTiO$_3$ under different
compressive in-plane strains. More specifically, for a given inplane lattice constant $a$,
both the out-of-plane lattice constant $c$ and atomic locations are
optimized by minimizing the total energy and forces. For the
optimization of atomic positions, residual forces on each atom
are required to be smaller than 10$^{-3}$ eV/{\AA}.
Effects of nuclei and core electrons are replaced by Troullier-Martins norm-conserving
type of pseudopotentials\citep{Troullier}; semi-core states Ti $3s$ and
$3p$ are treated as valence states for better accuracy.  Bloch wave functions in solid
are expanded using a mixed-basis set consisting of
numerical atomic pseudo-orbitals and plane waves.\citep{Fu_mix} A cutoff energy of
100 Ryd in wave-function expansion is used for both rotational and non-rotational SrTiO$_3$ on
equal footing, and has been checked to be sufficient.  Details on
generating pseudopotential, pseudo/all-electron matching radii, and
the accuracy checking were described in Ref.\citep{Vpseudo}.
Polarization structure, namely the contribution of a
given $\kpp$ string to the electronic polarization, is calculated
according to the equation\citep{King-Smith,Resta}
$    \phi(\kpp)={\rm Im}\{\ln\prod_{j=0}^{J-1}\,
    {\rm det}(\langle u_{m, k_j}| u_{n, k_{j+1}}\rangle)\}$,
where $J$ is the number of $\kpr$ points along the string. One
advantage of this equation over Eq.(\ref{EPhi_1})
is that $\phi(\kpp)$ here
does not depend on the arbitrariness of the phases of Bloch $u_{n\vek}$
wavefunctions at individual $\vek$ point.

\begin{figure}
\centering
\includegraphics[width=15cm]{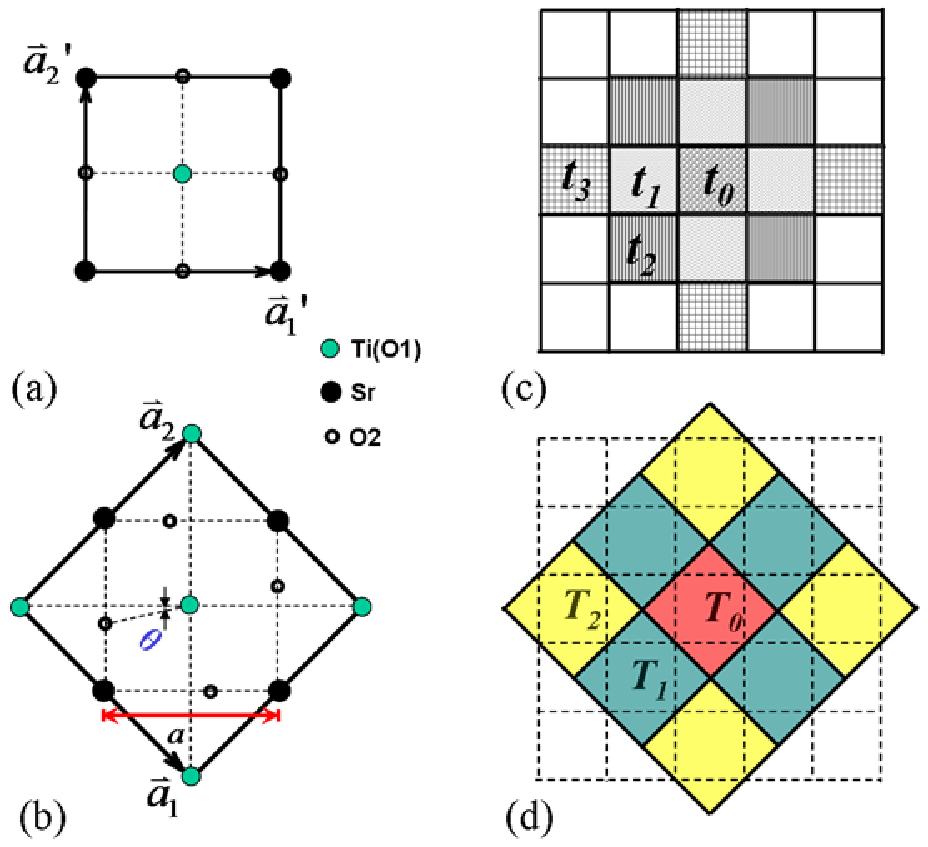}
\caption{(Color online) (a) Top view along the [001] direction of the 5-atom cell,
for SrTiO$_3$ without oxygen-octahedra rotation. (b) Top view of the 10-atom cell for
SrTiO$_3$ with oxygen rotation.  (c) Schematic illustration of  the interaction
between Wannier functions in different cells, for NR-SrTiO$_3$
consisting of 5-atom cells. (d) Similar as (c), but for
R-SrTiO$_3$ consisting of 10-atom cells.}
\label{struct}
\end{figure}

Figure \ref{struct}(a) and (b) show the top-view of the cell
structure for NR-SrTiO$_3$ without octahedral rotation of 5-atoms each cell,
and for R-SrTiO$_3$ with octahedral rotation of 10-atoms each cell,
respectively.
In NR-SrTiO$_3$, the lattice
vectors are $\vec{a}'_1$=$a\vec{i}$, $\vec{a}'_2$=$a\vec{j}$,
$\vec{a}'_3$=$c\vec{k}$. Biaxial compressive strains are applied in the
$\vec{a}'_1$ and $\vec{a}'_2$ directions. In R-SrTiO$_3$, the lattice vectors
are $\vec{a}_1$=$a(\vec{i}-\vec{j})$,
$\vec{a}_2$=$a(\vec{i}+\vec{j})$, $\vec{a}_3$=$c\vec{k}$. The
rotation angle $\theta$, as illustrated in Fig.\ref{struct}(b),
characterizes the deviation of the oxygen atoms from the non-rotational
positions in the $\vec{a}_1$-$\vec{a}_2$ plane. In the rest of
this section, we will briefly give our results on the
optimized structures for NR-SrTiO$_3$ and R-SrTiO$_3$ under different inplane
strains, since these data could be useful for experiments or other
theoretical calculations. The main results on polarization dispersion
will be discussed in the next section.

\begin{figure}
\centering
\includegraphics[width=15cm]{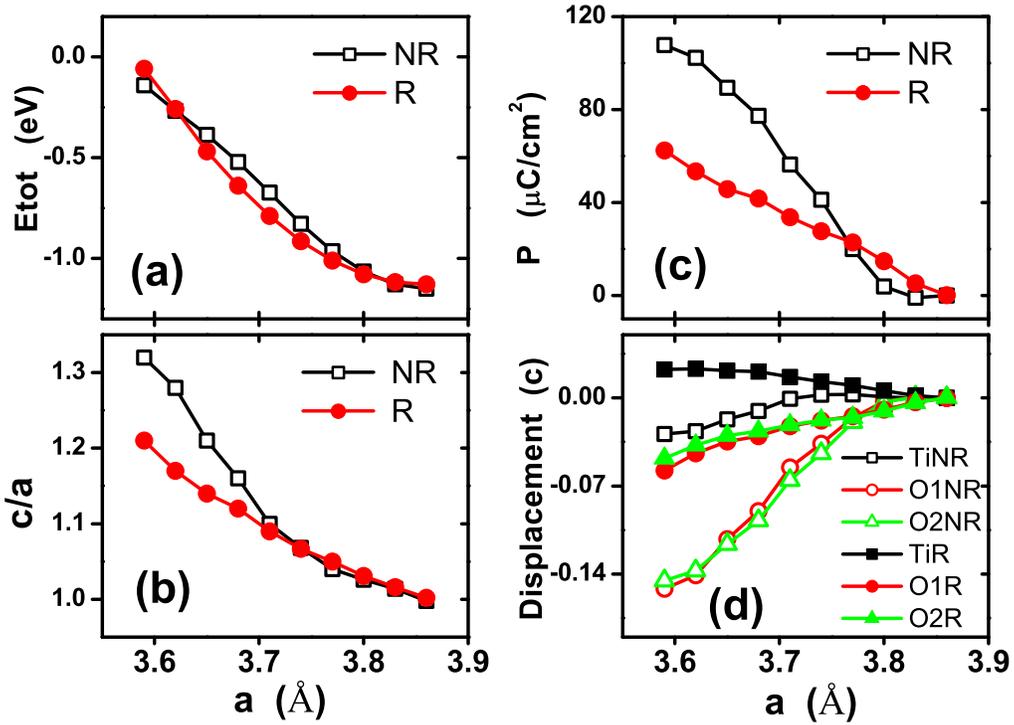}
\caption{(Color Online) The following calculated properties as a function of the in-plane $a$ lattice
constant in R-SrTiO$_3$ and NR-SrTiO$_3$: (a) total energy of a 10-atoms cell (with respect to a
basis of  -7474 eV), (b) the $c/a$ tetragonality, (c) the total (electronic
and ionic) polarization, (d) atomic displacements along the $c$-axis.
Atomic displacements in (d) are obtained by comparing the LDA-optimized atomic positions with
respect to the high-symmetric locations.  High-symmetric locations
are defined as: we first shift the Sr atom to the origin of the coordinate,
and then the high-symmetric atomic positions are $z=0$ for Sr and O1 atoms,
and $z=\frac{c}{2}$ for Ti and O2 atoms. O1 is apical while O2 is on
the base plane of an octahedra. } \label{pola}
\end{figure}

Figure \ref{pola}(a) gives the total energies per 10 atoms for NR-SrTiO$_3$
and R-SrTiO$_3$ at different in-plane $a$ lattice constants;
each curve is obtained by imposing symmetry constraint on the corresponding structure.
One main result in Fig.\ref{pola}(a) is that a crossover in
the ground-state structure occurs at $a_c$=3.62{\AA}. When $a>a_c$, R-SrTiO$_3$
has a lower energy than NR-SrTiO$_3$ and is thus more stable.
As $a$ is below $a_c$, NR-SrTiO$_3$ becomes more stable,
and the crystal of strained SrTiO$_3$ changes from rotational to non-rotational.
Meanwhile, we need to point out that at the
equilibrium lattice constant $a_0$=3.86{\AA}, our LDA simulations
yield a rather small energy difference between NR-SrTiO$_3$
and R-SrTiO$_3$, which is on the order of numerical uncertainty,
and does not allow us to tell which structure is more stable.
This, however, does not contradict with the fact that R-SrTiO$_3$ is
found more stable in experiments, since LDA calculations
can not capture the zero-point quantum fluctuation which is known
important in stabilizing the experimentally-observed rotational ground state of incipient SrTiO$_3$.
On the other hand, since the main purpose of this work is to understand the
polarization-structure difference between R-SrTiO$_3$ and NR-SrTiO$_3$,
the above uncertainty in the ground-state structure does not affect
our purpose of investigating the polarization structure.

The calculated $c/a$ tetragonality is given in Table \ref{caratio},
along with the rotation angle $\theta $ in R-SrTiO$_3$.
Under zero strain, both R-SrTiO$_3$ and NR-SrTiO$_3$ are found
to possess an equilibrium inplane lattice constant of $a_0$=3.86{\AA},
agreeing well with the experimental value\citep{LB} of 3.90{\AA} for R-SrTiO$_3$
at low temperature.
For R-SrTiO$_3$ at $a_0$, our theoretical rotation angle is
$\theta$=3.25$^0$, comparable to the experimental value\citep{Courtens} of
2.1$^0$ as well as 5.5$^0$ in another theoretical
calculation\citep{Sai_2}. Under inplane strains, the $c/a$ ratios
are shown in Fig.\ref{pola}(b).
One could see that by decreasing $a$,
$c/a$ increases for both NR-SrTiO$_3$ and R-SrTiO$_3$.
At small strains ($a$$>$3.71{\AA}), $c/a$'s of two structures remain close;
however, at large strains ($a$$<$3.71{\AA}), $c/a$ of NR-SrTiO$_3$
increases dramatically, and becomes evidently larger than that
in R-SrTiO$_3$. Interestingly, for the strain range studied here,
$c/a$ of R-SrTiO$_3$ depends on $a$ in a rather linear fashion. In contrast,
for NR-SrTiO$_3$, the $c/a$ ratio undergoes an obvious slope
change around $a$=3.71{\AA}. Strain-induced total polarization is shown
in Fig.\ref{pola}(c), where we see (i) for NR-SrTiO$_3$,
there is a critical inplane lattice constant $a\approx 3.82${\AA}, above
which polarization remains null; (ii) However, for R-SrTiO$_3$,
such a critical inplane lattice constant apparently does not
exist, and instead polarization increases almost
linearly with the decreasing inplane lattice constant.

The differences between NR-SrTiO$_3$ and R-SrTiO$_3$ in the $c/a$
tetragonality [Fig.\ref{pola}(b)] and in the electric polarization
[Fig.\ref{pola}(c)] can be explained microscopically by
the ferroelectric atomic off-center displacements, which are shown in Fig.\ref{pola}(d).
In NR-SrTiO$_3$, atomic displacements of O1 and O2 atoms are small when
$a$$>$3.80{\AA} and then increases drastically with a notable
slope change as $a$ is below 3.80{\AA}, which causes the sharp
rise of electric polarization in Fig.\ref{pola}(c). On the other hand,
in R-SrTiO$_3$ the off-center displacements of O1 and O2 vary
rather linearly with the decreasing inplane lattice constant, without significant
change in slope.  At $a$=3.59\AA, the
oxygen displacements in NR-SrTiO$_3$ are more than two times
as those in R-SrTiO$_3$.

\section{Polarization structure}

\subsection{Comparison on polarization dispersions}

\begin{figure}
\centering
\includegraphics[width=15cm]{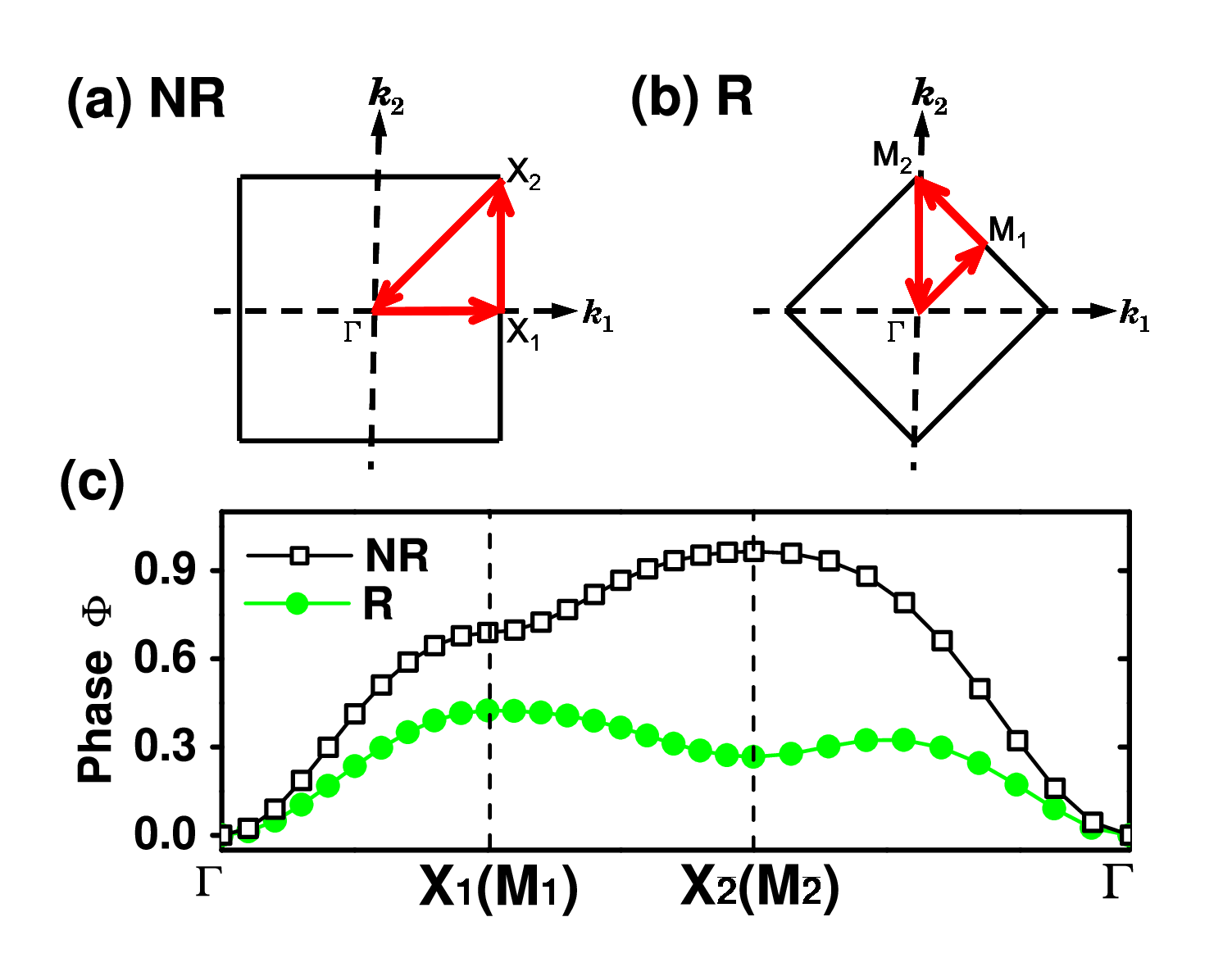}
\caption{(a) The two-dimensional Brillouin zone over which $\kpp$ is sampled, for
NR-SrTiO$_3$ with five atoms in one cell.
(b) The two-dimensional Brillouin zone over which $\kpp$ is sampled,
for R-SrTiO$_3$ with ten atoms in one cell. (c) Polarization
structures of NR-SrTiO$_3$ and R-SrTiO$_3$, both at $a$=3.71\AA.
For NR-SrTiO$_3$, 5-atom cell is used for the polarization structure
calculation.  For R-SrTiO$_3$, the horizontal $\kpp$ distance is rescaled by a factor
of $\sqrt{2}$  such that its dispersion structure
and the counterpart of NR-SrTiO$_3$ can be
plotted in the same figure.} \label{comp}
\end{figure}

We now present and discuss the polarization structures for SrTiO$_3$ with and without
oxygen-octahedra rotation, in particular the similarity and difference between them.
Under zero strain, both NR-SrTiO$_3$
and R-SrTiO$_3$ have null polarizations, and the $\phi (\kpp)$ phase
is trivial and vanishes for all $\kpp$ points.
On the other hand, under compressive strains, both systems develop electric polarizations.
In the following, we chose an inplane lattice constant $a$=3.71{\AA},
and determine the polarization structure for NR-SrTiO$_3$ and R-SrTiO$_3$
by using the optimized atomic positions and cell shapes of each system.
Since NR-SrTiO$_3$ and R-SrTiO$_3$ have different sizes of unit cells,
their Brillouin zones (BZ) are illustrated in Fig.\ref{comp}(a) and (b), respectively.
The triangles highlighted by thick lines in Fig.\ref{comp}(a) and (b)
are the high-symmetry paths along which the $\kpp$-dependent
polarization structure is examined. The calculated $\phi(\kpp)$ phase
structures are shown in Fig.\ref{comp}(c).
To facilitate comparison, both polarization structures are shifted
to make $\phi(\Gamma)=0$ at $\Gamma$; meanwhile,
the horizontal axis of $\kpp$ distance in R-SrTiO$_3$ is rescaled
by a factor of $\sqrt{2}$ in Fig.\ref{comp}(c), such that the two
polarization dispersions can be given in the same figure.

Inspection of the polarization structures in Fig.\ref{comp}(c) reveals:
(i) The dispersion bandwidth in NR-SrTiO$_3$ ($\sim$0.9) is
considerably larger than that in R-SrTiO$_3$ ($\sim$0.4).
(ii) In NR-SrTiO$_3$, the largest $\phi(\kpp)$ contribution comes from
the $X_2$ point at the zone corner, while in R-SrTiO$_3$ the largest
contribution is at $M_1$ which is on the zone boundary. (iii) In NR-SrTiO$_3$, $X_1$
point is a saddle point, while in R-SrTiO$_3$ there is no saddle point.
Furthermore, in R-SrTiO$_3$, $M_2$ is a local minimum
and another local maximum appears along the $M_2-\Gamma$ line,
which is not the case in NR-SrTiO$_3$.  As a
result, the two polarization dispersion curves have rather
different curvatures.

While the above differences in the polarization structure between
NR-SrTiO$_3$ and R-SrTiO$_3$ are interesting, they are also puzzling.
With some effort, we find two possible reasons that might be responsible for these differences.
The first one is folding effect. It is known in band-structure calculations
that when one changes the size of unit cell, there is band-folding effect.
However, it is unclear whether the folding effect also exists for polarization
dispersion. If the answer is yes, one may also desire to
establish a rigorous proof on the existence of such an effect, and better yet,
to uncover the precise manner on how the $\phi(\kpp)$ dispersion curve folds itself.
The second possible reason is rotational effect.
When octahedral rotation is allowed, the atom-atom interaction in R-SrTiO$_3$
is anticipated to be different from in NR-SrTiO$_3$, which could alter the electron distribution
and thus the polarization structure. In the following, we will determine and analyze the
consequences of both effects.

\subsection{Folding effect}

To determine the significance of each of the
above two effects (i.e., folding effect and rotational effect),
it is preferable to design an approach that is able to investigate each effect separately.
We first investigate the folding effect, by conducting
the following calculations in order to avoid deliberately the rotational effect
so that the latter will not play a role.
We begin with an LDA-optimized 5-atom cell
for NR-SrTiO$_3$ at a given inplane lattice (the solid of which is to be
denoted as NR5), and then double the size of cell and
convert it into a 10-atom cell (the solid
of which is to be denoted as NR10, once again without oxygen rotation).
The atomic positions and the electron charge
densities in NR5 before conversion and in
NR10 after conversion are thus equivalent. Also note that
there is no oxygen rotation in either NR5 or NR10;
differences in polarization structure between these two solids
thus come from the folding effect alone.

\begin{figure}
\centering
\includegraphics[width=15cm]{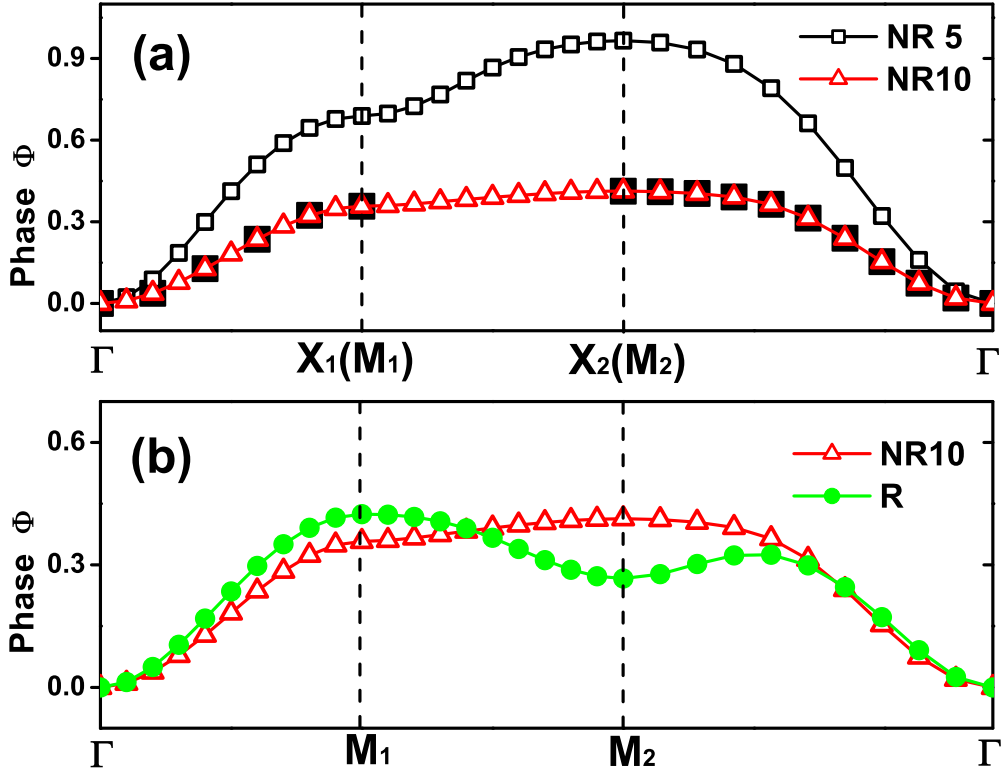}
\caption{(a) Polarization structures of NR5 and NR10 at
inplane lattice constant $a$=3.71{\AA}. The analytic
result using Eq.(\ref{Efold}) is also shown as solid squares.
(b) Comparison between the polarization structure of NR10 and
that of R-SrTiO$_3$ at $a$=3.71{\AA}.} \label{cellsize}
\end{figure}

Fig.\ref{cellsize}(a) presents the polarization structures of NR5
and NR10, both at $a$=3.71{\AA}. One sees that the polarization structure of
NR10 (empty triangles in Fig.\ref{cellsize}a) appears
very different from that of NR5 (empty squares).
In particular, (i) The bandwidth of the NR5 curve is about
twice as large as the NR10 case. (ii) For NR5,
$X_2$ is the maximum, and $X_1$ is a saddle point as we saw
previously. However for NR10,
$M_1$ and $M_2$ have similar $\phi({\kpp})$ values, making
the curve between $M_1$ and $M_2$ evidently flat.
(iii) The curvatures are different in two solids.
As a result, apparently no relationship could be
found for the two polarization-dispersion curves at first glance.
More disturbingly, one may also wonder whether the polarization structure is uniquely
defined for a given crystal, or whether it is meaningful to
make a sensible comparison on polarization dispersion.
To answer this, we will establish a rigorous relationship
between the two curves in Fig.\ref{cellsize}(a),
using an analytical approach based on Wannier functions.

We start with the expression for the polarization phase
structure \citep{Yao_Pol},
$\phi(\kpp)=\frac{2\pi}{c}\sum_{\Rpp}\sum_{n=1}^M \int \rpr
W_n^*(\vec{r})W_n(\vec{r}-\Rpp) e^{i\kpp \cdot \Rpp}d\vec{r},$
where $W_n(\ver)$ is the Wannier function of the solid defined as
$W_n(\ver-\veR)=\frac{\sqrt{N}\Omega}{(2\pi)^3}
\int_{BZ}d\vec{k}e^{i\vec{k}\cdot(\ver-\veR)}u_{nk}(\ver) $.
Subscript $\parallel$ (or $\perp$) in the equation
means the projection component parallel (or perpendicular, respectively) to the
direction of the considered polarization; $\Rpp$ thus refers to a lattice vector
perpendicular to the direction of the polarization. $M$ is the number
of occupied bands in solid. By defining parameters
\begin{equation}
t(\Rpp)=\frac{2\pi}{c}\sum_{n=1}^M\int \rpr
W^*_n(\ver)W_n(\ver-\Rpp)d\ver \  , \label{EP_t}
\end{equation}
we then arrive at a concise and analytic expression for polarization structure,
\begin{equation}
\phi(\kpp)=\sum_{\Rpp}t(\Rpp)e^{i\kpp \cdot \Rpp} \ . \label{EP_TB}
\end{equation}
Sum over $\Rpp$ in Eq.(\ref{EP_TB}) should in principle been carried
out to infinite distance. In practice, owing to the localization nature of
Wannier functions in dielectric materials\citep{Kohn_WF,Marzari},
it is often sufficient to truncate $\Rpp$ within a few near neighbors (NNs).

To simplify the formulations, we will consider, for
NR5, up to the 3rd nearest-neighbor interactions in Eq.(\ref{EP_TB}),
with parameters $t_0$, $t_1$, $t_2$, and $t_3$ representing respectively
the on-site, first NN (1NN), second NN (2NN), and third NN (3NN) Wannier-function interactions as
schematically illustrated in Fig.\ref{struct}(c).
On an equal footing, the same truncation distance
is used for Wannier-function interactions in NR10 where cell size is doubled.
However, we should point out that our final conclusion on folding effect
is generally applicable and does not depend
on the distance within which Wannier-function interactions are truncated.
Furthermore, to avoid ambiguity, we will use $\Rpp'$ and $\phi'$ for
NR5, while we use $\Rpp$ and $\phi$ for NR10.
For NR5 within 3NN truncation, $\Rpp'$s are $\Rpp'$=(0,0) (on site);
($\pm a$,0) and (0,$\pm a$) (1NNs); ($\pm a$,$\pm a$) (2NNs);
($\pm 2a$,0) and (0,$\pm 2a$) (3NNs).
With tetragonal symmetry, we can rewrite Eq.(\ref{EP_TB}) as
\begin{equation}
 \begin{split}
  \phi'(\kpp)=t_0+
2t_1[\cos(k_1a) + \cos(k_2a)] \\
+4t_2\cos(k_1a)\cos(k_2a)
+2t_3[\cos(2k_1a)+\cos(2k_2a)]\ ,
 \end{split}
 \label{Ephi5}
\end{equation}
where $k_1$ and $k_2$ are the components of
$\kpp$, namely $\kpp=(k_1,k_2)$.

For NR10, there are two sets of Wannier functions in one unit cell,
one set being the same as those in NR5 and another set having
identical shapes but displaced by the lattice vector $\vec{a}'_1$ of NR5.
Meanwhile, we denote the Wannier-interaction parameters in NR10 as $T_i$
as schematically shown in Fig.\ref{struct}(d), to be distinguished from
$t_i$ in NR5. Within the same truncation distance as in NR5, we only
need to consider in NR10 the interactions between Wannier functions up to 2NNs,
where $\Rpp $=(0,0) (on site); ($\pm a$,$\pm a$) (1NNs);
($\pm 2a$,0) and (0,$\pm 2a$) (2NNs). The polarization structure in NR10
could be determined as
\begin{equation}
 \begin{split}
  \phi(\kpp)=T_0
+ 4T_1\cos(k_1a)\cos(k_2a)\\
+2T_2[\cos(2k_1a)+\cos(2k_2a)]\ .
 \end{split}
 \label{Ephi10}
\end{equation}
By the definition of Eq.(\ref{EP_t}), it can be further shown that
$t_i$s in NR5 and $T_i$s in NR10 are related by
$T_0=2t_0$, $T_1=2t_2$, and $T_2=2t_3$, since the Wannier
functions are identical in both systems. Combining
Eq.(\ref{Ephi5}) with Eq.(\ref{Ephi10}), one can find the
following relation between $\phi$ and $\phi'$ phases
\begin{equation}
\phi(\kpp)=\phi'(\kpp)+\phi'(\kpp - \vec{b}_2)\  ,
\label{Efold}
\end{equation}
where $\vec{b}_2=\frac{\pi}{a}(\vec{i}+\vec{j})$ is one of the
reciprocal lattice vectors of NR10. In other words,
for an arbitrary $\kpp$ point within the triangle
$\Gamma$M$_1$M$_2$ in the irreducible BZ of NR10, the phase
angle $\phi(\kpp)$ of NR10 could be determined from
the $\phi'$ phases of NR5, at two
$\kpp$ points within the triangle $\Gamma$X$_1$X$_2$ in the
irreducible BZ---one contribution being at the same $\kpp$ point
(the first term in Eq.\ref{Efold}) and the other is the
folded contribution at the $\kpp-\vec{b}_2$ point
(the second term in Eq.\ref{Efold}).
We thus prove that the folding effect indeed exists for the
polarization dispersion structure.

We next would like to numerically confirm the validity of Eq.(\ref{Efold}).
We apply Eq.(\ref{Efold}) and use the right-hand side of this equation
to determine the polarization structure $\phi(\kpp)$ of NR10.
The results are given as solid squares in Fig.\ref{cellsize}(a),
where we see that the analytic results of Eq.(\ref{Efold}) coincide with
those LDA calculated results of NR10.  This
demonstrates that the polarization structures of NR5 and NR10, although quite
different looking, are equivalent and are thus physically
meaningful.

\subsection{Rotational Effect}

After revealing the existence of folding effect, we proceed to
investigate rotational effect. To single out the rotation effect, we compare
in Fig.\ref{cellsize}(b) the polarization structures of NR10
(without rotation) with that of R-SrTiO$_3$ (with rotation), both
at the inplane lattice constant $a$=3.71{\AA}.
Note that both systems have 10 atoms in one unit cell, and
the folding effect is irrelevant here.

After eliminating the folding effect, we see in Fig.\ref{cellsize}(b)
that the bandwidths of the polarization structures with and without
rotation become notably similar.
However, the two polarization structures still show important
differences: (i) $\phi(M_2)$ is a global maximum in non-rotational
NR10, whereas being a local minimum in rotational structure;
(ii) $\phi(M_1)$ is still a saddle point in NR10 as in NR5,
but is a global maximum in rotational case;
(iii) Between $M_1$ and $M_2$, very little dispersion
exists for the polarization structure of NR10,
which is not the case for R-SrTiO$_3$ where sizeable dispersion
is evident;  (iv) A local maximum is observed along $M_2-\Gamma$
in R-SrTiO$_3$, and such a maximum does not exist in
the case of NR10.
Rotation of the oxygen octahedra alters the interactions among
atoms (also the Wannier functions), which is responsible for the
above-described differences in polarization structure.
Differences (i)-(iv) also reveal that polarization
structure is sensitive to the subtle changes in atomic positions and
crystal structure, and could be used as an effective tool for studying
material properties.

\subsection{Response of polarization structure to in-plane strain}

\begin{figure}
\centering
\includegraphics[width=15cm]{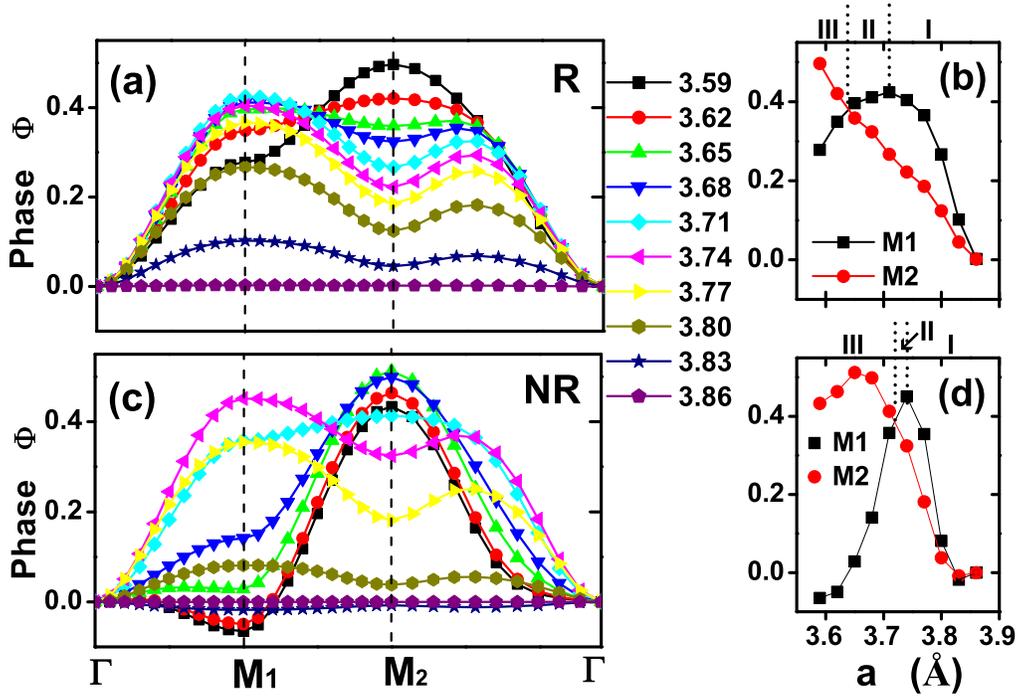}
\caption{(Color online) (a) Polarization structure of R-SrTiO$_3$ at different in-plane
lattice constants; (b) Phase angles at $M_1$ and $M_2$ as a function of
the in-plane $a$ lattice constant, for R-SrTiO$_3$.
(c) Polarization structure of non-rotational NR10 at different
in-plane lattice constants; (d) Phase angles at $M_1$ and $M_2$ as
a function of the in-plane $a$ lattice constant, for NR10.
All polarization dispersion curves are rigidly shifted such that
$\phi(\Gamma)=0$ for the sake of comparison.} \label{m1m2}
\end{figure}

Compressive inplane strains---caused by lattice mismatch between epitaxially
grown solid and substrate---modify interatomic interactions
in an anisotropic fashion, by strengthening the interaction
along inplane directions but weakening the interaction along
the out-of-plane direction. This modification affects the
structural and electric properties.  As a consequence of the altered
interatomic interaction, the polarization structure might
also change accordingly. In this part, we examine the strain dependence of
the polarization structure in R-SrTiO$_3$ with octahedral rotation, and
contrast it with that in the non-rotational structure.
To eliminate the folding effect,
we use a unit cell of 10 atoms for non-rotational SrTiO$_3$ (i.e., NR10),
same as for R-SrTiO$_3$.
The $\kpp$ points are along the $\Gamma\rightarrow M_1\rightarrow M_2\rightarrow
\Gamma$ path as in Fig.\ref{comp}(b). The calculated
polarization structures at different in-plane lattice constants
are shown in Fig.\ref{m1m2}. More specifically, for R-SrTiO$_3$, its polarization
dispersions under different inplane $a$ lattice constants are shown
in Fig.\ref{m1m2}(a), while Fig.\ref{m1m2}(b) summarizes the phase $\phi(\kpp)$ angles
at high-symmetric $M_1$ and $M_2$ points as a function of the inplane
lattice constant.
For non-rotational NR10 structure, the polarization dispersions are given in
Fig.\ref{m1m2}(c) while phase angles at $M_1$ and $M_2$ are
summarized in Fig.\ref{m1m2}(d).

Let us focus on R-SrTiO$_3$ first. In Fig.\ref{m1m2}(a), we see
that with the increase of in-plane strain, the polarization
structures change as following: (i) At zero strain or
$a$=3.86{\AA}, the polarization $\phi(\kpp)$ phase vanishes at all
$\kpp$ points, as a result of the existing inversion symmetry;
(ii) Upon application of small strain, the bandwidth of
polarization structure begins to increase. For $3.83{\rm \AA}\geq
a\geq 3.71{\rm \AA}$, the $\phi(\kpp)$ angle peaks at $M_1$, while
being a local minimum at $M_2$.  Meanwhile, a local maximum is
observed along $M_2$-$\Gamma$. On the other hand, the zone center,
$\Gamma$, contributes the least to the total electronic
polarization; (iii) When $3.71{\rm \AA}\geq a\geq 3.65{\rm \AA}$,
$\phi(M_2)$ increases while $\phi(M_1)$ slightly decreases, making
the $\phi(\kpp)$ curve of $a$=3.65{\AA} notably flat between
$M_1$ and $M_2$. (iv) As $a$ is further decreased below 3.65{\AA},
$M_2$ turns into the maximum point, while $M_1$ becomes a saddle
point. Meanwhile, the local maximum previously existing along
$M_2$-$\Gamma$ disappears. Quantitatively, evolutions of the
$\phi(\kpp)$ phases at $M_1$ and $M_2$ with the inplane
lattice constant are better visualized in Fig.\ref{m1m2}(b). The
evolution can be separated into three regions: Region I where
$\phi(M_1)$ and $\phi(M_2)$ both increase with the increasing
strain, Region II where $\phi(M_1)$ starts to decline but
$\phi(M_1)$ still contributes greater than $\phi(M_2)$, and Region
III where crossover between $\phi(M_1)$ and $\phi(M_2)$ occurs and
$\phi(M_2)$ now becomes the maximum contribution.

The differences between R-SrTiO$_3$ and NR10, in their
responses of polarization structure to inplane strain, can be
seen by comparing Fig.\ref{m1m2}(d) with Fig.\ref{m1m2}(b),
which yields the following observations: (i) When $a$ is near 3.83{\AA}, $\phi(M_1)$ and
$\phi(M_2)$ both increase linearly in R-SrTiO$_3$
[Fig.\ref{m1m2}(b)], while they remain nearly zero in NR10
[Fig.\ref{m1m2}(d)]. It reveals once again a lack of critical
inplane strain in R-SrTiO$_3$. However, in NR10 without
rotation, $a$ need be below 3.83{\AA} in order to produce
sizable polarization. (ii) When $3.83${\AA}$\leq a\leq3.74${\AA},
$\phi(M_2)$ increases quickly with strain both in R-SrTiO$_3$ and
in NR10, but the increasing slope in NR10
[Fig.\ref{m1m2}(d)] is much larger than in R-SrTiO$_3$
[Fig.\ref{m1m2}(b)]. During this region of strain, the
polarization in NR10 is more sensitive to the inplane strain with
larger piezoelectric response than in R-SrTiO$_3$, and
microscopically $\phi(M_2)$ is largely responsible. (iii) Although
the responses in Fig.\ref{m1m2}(d) can also be divided into three
stages I, II and III for NR10, the inplane lattice constants
delimiting the boundary of different stages differ, however, from
those in Fig.\ref{m1m2}(b). (iv) In NR10, $\phi(M_1)$ declines
sharply at stages II and III, and as a result, $\phi(M_2)$ becomes
dominating when $a\le3.68${\AA}. In R-SrTiO$_3$, $\phi(M_1)$
declines much slowly at stages II and III. (v) $\phi(M_2)$ in NR10
reaches its maximum around $a$=3.65{\AA}, then starts to decrease. In
R-SrTiO$_3$, $\phi(M_2)$ nevertheless keeps increasing.

Response of polarization structure to inplane strain is thus
found interestingly different in R-SrTiO$_3$ with oxygen
rotation as compared to that in NR10 without rotation. These
differences should, on the microscopic level, originate from the
different interaction among the Wannier functions in two
structures. In the following, we will extract from the
polarization structure the interactions among Wannier functions.

\subsection{Implication on interaction
between Wannier functions}

\begin{figure}
\centering
\includegraphics[width=15cm]{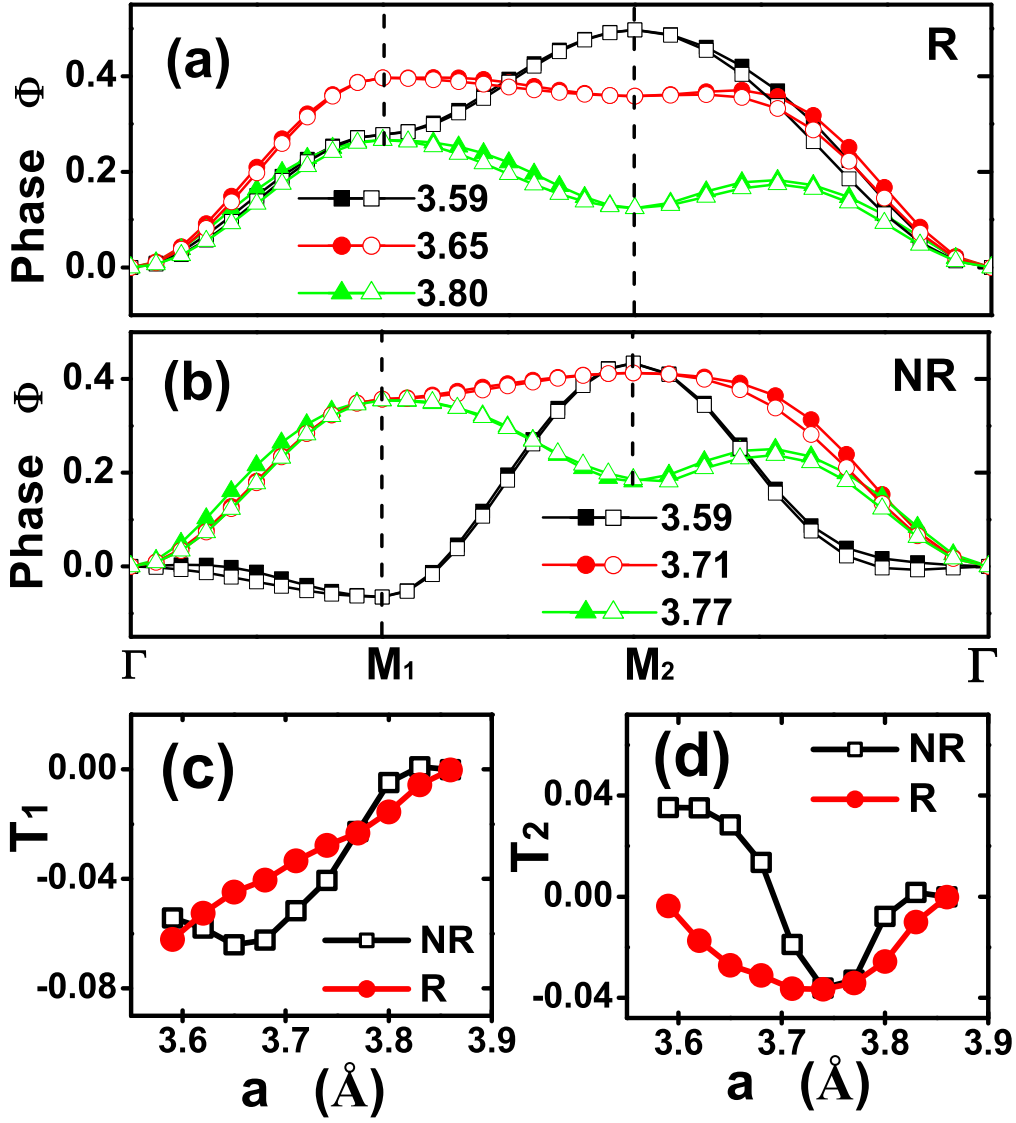}
\caption{(a) Analytic polarization dispersions (open symbols) determined
from Eq.(\ref{Ephi10}), as compared to the direct results (filled symbols)
calculated from the modern theory of polarization, for R-SrTiO$_3$
at inplane lattice constants $a$=3.80, 3.65, and 3.59{\AA}.
Lines are guides for eyes. Since analytic results agree well with direct
calculation results, most of open symbols are on top of filled symbols.
(b) Similar as (a), but for NR10 at $a$=3.77, 3.71, and 3.59{\AA}.
(c) Dependence of the $T_1$
quantity on the inplane $a$ lattice constant, for R-SrTiO$_3$
and NR10. (d) Dependence of the $T_2$ quantity on the
inplane lattice constant.}
\label{Ffit}
\end{figure}

Our procedure of determining Wannier-function interaction begins with
Eq.(\ref{Ephi10}), which was derived for crystals
consisting of 10 atoms in one unit cell, and is thus applicable
to both R-SrTiO$_3$ and NR10 though with different interaction parameters.
Applying Eq.(\ref{Ephi10}) to $\phi(\kpp)$ at special $\kpp$ points
at $\Gamma$, $M_1$ and $M_2$, we obtain
$\phi(M_1)-\phi(\Gamma)=-4T_1-8T_2$ and
$\phi(M_2)-\phi(\Gamma)=-8T_1$.
From $\phi(M_1)$ and $\phi(M_2)$, one can thus determine $T_1$ and
$T_2$ quantities. As shown in Fig.\ref{struct}(d),
$T_1$ (or $T_2$) corresponds to the Wannier-function
interaction between a given cell with the 1st (or 2nd) neighboring
cell, manifesting interatomic interaction on microscopic level.
The obtained $T_1$ and $T_2$ quantities are then used in Eq.(\ref{Ephi10})
to determine analytically the whole polarization dispersion curve.
Note that only $\phi$ phases at $M_1$ and $M_2$ points are fitted.
To evaluate the validity of our analytic formulas,
we choose three inplane lattice constants
across the entire strain range considered in this work,
and plot the {\it analytic} polarization dispersions as
open symbols in Fig.\ref{Ffit}(a) for R-SrTiO$_3$ and in
Fig.\ref{Ffit}(b) for NR10.  One sees that for each inplane
lattice constant, the polarization structures analytically determined
from Eq.(\ref{Ephi10}), and the direct results obtained from
the Berry-phase modern theory of polarization, agree well. More specifically,
for the $a$=3.80{\AA} curve in Fig.\ref{Ffit}(a),
the small dispersion hump between $M_2$ and $\Gamma$
is well reproduced by analytical result.
Furthermore, at $a$=3.59{\AA} which is a more stringent test on the
analytic formulation with truncated interaction,
the saddle-point nature at $M_1$ in Fig.\ref{Ffit}(a),
as well as the subtle minimum at $M_1$ in Fig.\ref{Ffit}(b), are
also predicted by the analytic expressions. These demonstrate the
validity of our analytic formulation; the obtained $T_1$ and
$T_2$ values are thus reliable.

The obtained $T_1$ and $T_2$ quantities are given in Table \ref{T1T2} and
shown in Fig.\ref{Ffit}(c) and (d). One can see from
Fig.\ref{Ffit}(c) and (d) that, for a given inplane
lattice constant, $T_i$ is notably
different in R-SrTiO$_3$ with respect to that in non-rotational
NR10. For instance, at $a$=3.68{\AA}, $T_1$ in NR-SrTiO$_3$ is $\sim$50\% larger
than in R-SrTiO$_3$, and $T_2$ in NR-SrTiO$_3$ even has a different sign
from $T_2$ in R-SrTiO$_3$. These quantitative differences in
microscopic interactions indicate that polarization structure is
indeed a rather powerful tool to understand physical properties.
We also note that, at large inplane strain such as $a$=3.59{\AA},
$T_2$ is large in NR-SrTiO$_3$ but nearly vanishes in R-SrTiO$_3$
[Fig.\ref{Ffit}(d)], implying that interactions in rotational and
non-rotational structures are significantly different. Furthermore, we observe that
at very small inplane lattice constants near $a$=3.59{\AA}, $T_1$ and $T_2$ tend to saturate
in NR-SrTiO$_3$ but not in R-SrTiO$_3$.

\subsection{Determination of higher-order Wannier-function interaction}

One relevant question in the theory of polarization structure is
the determination of the $t_i$s quantities,
particularly those higher-order $t_i$s that correspond to larger $\Rpp$s,
since by knowing $t_i$s the $\phi(\kpp)$ phases can be readily obtained
via Eq.(\ref{EP_TB}).
The $t_i$ quantities connect the microscopic interaction with macroscopic
$\kpp$-dependent polarization structure. As one useful
outcome of the current study, we describe an approach that can be
utilized to determine the higher-order $t_i$ interactions.

\begin{figure}
\centering
\includegraphics[width=15cm]{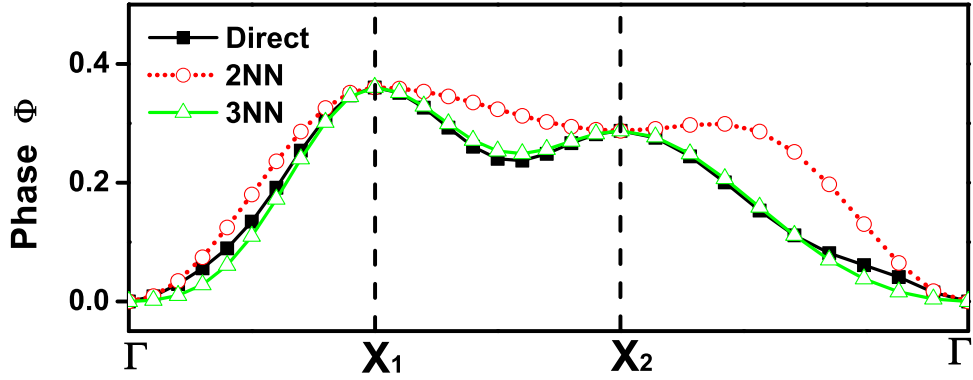}
\caption{Polarization structure of NR-SrTiO$_3$ at $a$=3.59{\AA}
computed from direct Berry Phase calculation (solid squares),
2NN approximation (open circles and dotted line), and 3NN approximation (open triangles).
Lines are guides for eyes.} \label{F3NN}
\end{figure}

The approach works in general. But for the sake of
convenience in our discussion, we will take NR-SrTiO$_3$ (without oxygen rotation)
of $a$=3.59{\AA} as an example. For NR-SrTiO$_3$,
the smallest unit cell consists of 5 atoms, and the polarization
structure directly calculated from Berry phase is given in Fig.\ref{F3NN}.
Now if we use a lower-order 2NN approximation in Eq.(\ref{EP_TB}),
with interactions $t_0$, $t_1$ and $t_2$ in Fig.\ref{struct}(c), to analytically
mimic the polarization structure, it is easy to derive that
the polarization structure will be
$\phi(\kpp)=t_0+
2t_1[\cos(k_1a) + \cos(k_2a)]
+ 2t_2[\cos(k_1+k_2)a + \cos(k_1-k_2)a ]$.
By using the phase angles at $\Gamma$,
$X_1$ and $X_2$ points, one can determine $t_0$, $t_1$ and $t_2$.
The polarization dispersion within the 2NN approximation
can be subsequently obtained and is given in Fig.\ref{F3NN}.
One sees that the agreement between the directly calculated polarization curve
and the 2NN-approximated curve is not good. For instance,
$\phi(X_2)$ is incorrectly predicted by the 2NN approximation to be a local minimum,
instead of a local maximum as by the direct calculation. Meanwhile,
the dispersion curvatures along $X_1$-$X_2$ and along $X_2-\Gamma$
are wrong. We thus see that higher-order interactions among Wannier
functions are needed. This is not surprising, since at
small inplane lattice constants, interactions between farther neighboring cells
remain strong and can not be neglected.

To determine higher-order $t_i$ quantities, one possible approach
(denoted as approach I) is to use higher-order $\phi(\kpp)$ expression
such as Eq.(\ref{Ephi5}) within the 3NN approximation,
and in Eq.(\ref{Ephi5}) four $t_i$s need be fitted. In this approach,
if even higher-order approximations than 3NN are used, more $t_i$s have to be
fitted at many $\kpp$ points, which becomes increasingly complex
and less trackable.
An alternative approach (denoted as approach II) is the following.
First, we recognize from the finding of the
folding effect in Sec.IIIB that, when one doubles the size of
unit cell from 5-atoms to 10-atoms, the solid is the same, but the
$t_1$ term drops as seen by comparing Eq.(\ref{Ephi10}) with Eq.(\ref{Ephi5}).
As a result, one is still left with three (not four) parameters,
and can be determined by fitting $\phi(\kpp)$s at three $\kpp$ points.
Similarly, one can determine even higher-order Wannier-function interactions
by varying the size of unit cell. In fact, when one increases
the size of unit cell (doubling, tripling, etc), the
interactions between those Wannier functions---that are located in different cells
of previously smaller cell size but fall in the same cell of the larger cell size---are dropped.
As a result, with the large unit cell, determination of new
Wannier-function interactions becomes simplified,
which corresponds to higher-order NN interactions with respect to the previous
smaller unit cell. For instance, in Eq.(\ref{Ephi10}) when cell size is doubled, a new interaction
parameter $T_2$ (i.e., $t_3$ since $T_2=2t_3$) is determined.
This approach is, in spirit, similar to the renormalization-group (RG) approach used
in understanding critical behaviors of phase transition \citep{Wilson}.
Of course, the benefit of approach II is at the expense of computing
the Berry-phase polarization structure in larger unit cell, but the
increase of computation is rather small.

To examine how approach II works, we give in Fig.\ref{F3NN} the
analytical dispersion determined within the 3NN approximation,
with $t_1$ obtained from 5-atom-cell calculation, and with $t_2$
and $t_3$ obtained from 10-atom-cell calculation. The obtained
values are $t_1$=-3.590$\times10^{-2}$,
$t_2$=-2.711$\times10^{-2}$, and $t_3$=1.761$\times10^{-2}$. As
one could easily see from Fig.\ref{F3NN}, inclusion of $t_3$
interaction in the 3NN approximation yields a much better
dispersion than the 2NN one. The local-maximum nature at $X_2$,
the characteristic dip between $X_1$ and $X_2$, and the curvature
between $X_2$ and $\Gamma$ in the direct Berry-phase calculation
results are well described by the 3NN approximation but not the
2NN approximation.

\section{summary}

Polarization structure in R-SrTiO$_3$ with rotational instability
is studied, and contrasted with the dispersion in NR-SrTiO$_3$
where oxygen-octahedra rotation is deliberately not allowed by symmetry
constraint. Influences of in-plane compressive strains on the
polarization dispersion as well as on the phase contributions at
high-symmetric $\kpp$ points are investigated. Microscopic
interactions are deduced by means of Wannier-function formulation,
and the dependencies of $T_1$ and $T_2$ interactions on the
inplane strain are determined.

Our specific findings are summarized as following: (i) The
polarization structures of ST with and without oxygen rotation, at
the same in-plane lattice constant, are rather different. These
differences could be attributed to two reasons, the folding effect
and the rotational effect. (ii) We have proved, both numerically
and analytically via the Wannier-function formulation, that
folding effect exists in polarization structure. As a result,
polarization structures are equivalent for differently chosen unit
cells, and are thus physically meaningful. We also reveal how
polarization structure folds itself. (iii) After elimination of
folding effect by choosing the same unit cell, the polarization
structures for rotational and non-rotational geometries have
similar bandwidth. But the dispersion curvatures and individual $\phi(\kpp)$ phases are still
considerably different as shown in Fig.\ref{cellsize}(b), which
reveals that polarization structure is sensitive to the structure
distortion and is indeed an effective tool to study materials
properties of dielectrics. (iv) With application of in-plane
compressive strain, the $\phi(\kpp)$ phase at high-symmetric $M_1$
point is found to decline sharply at stage III in NR-SrTiO$_3$
[Fig.\ref{m1m2}(d)], and in contrast, the decline is much slower in R-SrTiO$_3$
[Fig.\ref{m1m2}(c)]. Furthermore, in NR-SrTiO$_3$, the
$\phi(\kpp)$ phase at another high-symmetric $M_2$ point first increases at small strain and
then declines at high strain. However, in R-SrTiO$_3$, the $\phi$
phase at $M_2$ keeps increasing throughout the considered strain
range. (v) From polarization structure, we obtained the
microscopic $T_1$ and $T_2$ interactions as given in
Fig.\ref{Ffit}(c) and (d). $T_i$ interaction is found to differ
significantly in R-SrTiO$_3$ and NR-SrTiO$_3$. Moreover, $T_2$ in
Fig.\ref{Ffit}(d) is revealed to exhibit an interesting
non-monotonous dependence on inplane lattice constant. (vi) We have
further described a renormalization-like method to determine
higher-order Wannier-function interactions by varying the size of
unit cell. We revealed that the 3NN approximation is necessary for
an accurate description of the polarization structure of
NR-SrTiO$_3$ at $a$=3.59{\AA}.

\section{Acknowledgement}

This work is supported by the Office of Naval Research.

\newpage

\begin{table}
\caption{The LDA-optimized $c/a$ ratios for NR-SrTiO$_3$ and
R-SrTiO$_3$ under different inplane lattice constants. For R-SrTiO$_3$,
the calculated rotation angle is also given.} \label{caratio}
\begin{tabular}{c|c|cc}
  \hline\hline
  & NR-SrTiO$_3$  & \multicolumn{2}{c}{R-SrTiO$_3$} \\
  $a$({\AA}) & $c/a$ & $c/a$ & angle $\theta (^o)$ \\
  \hline
  3.59 & 1.32 & 1.21 & 12.70 \\
  3.62 & 1.28 & 1.17 & 12.31 \\
  3.65 & 1.21 & 1.14 & 11.71 \\
  3.68 & 1.16 & 1.12 & 10.92 \\
  3.71 & 1.10 & 1.09 & 9.95 \\
  3.74 & 1.068 & 1.067 & 8.80 \\
  3.77 & 1.040 & 1.050 & 7.81 \\
  3.80 & 1.026 & 1.031 & 6.57 \\
  3.83 & 1.014 & 1.016 & 5.39 \\
  3.86 & 0.998 & 1.002 & 3.25 \\
  \hline\hline
\end{tabular}
\end{table}

\begin{table}
\caption{$T_1$ and $T_2$ quantities (in radian) at different inplane lattice
constants. } \label{T1T2}
\begin{tabular}{c|cc|cc}
  \hline\hline
  & NR10 &  & R-SrTiO$_3$ &  \\
  $a$ ({\AA}) & $T_1$ ($10^{-2}$) & $T_2$ ($10^{-2}$) & $T_1$ ($10^{-2}$) &
  $T_2$($10^{-2}$) \\
  \hline
  3.59 & -5.422 & 3.522 & -6.210 & -0.372 \\
  3.62 & -5.798 & 3.512 & -5.256 & -1.736 \\
  3.65 & -6.408 & 2.842 & -4.848 & -2.714 \\
  3.68 & -6.234 & 1.350 & -4.064 & -3.116 \\
  3.71 & -5.162 & -1.888 & -3.334 & -3.636 \\
  3.74 & -4.058 & -3.612 & -2.780 & -3.662 \\
  3.77 & -2.272 & -3.298 & -2.322 & -3.412 \\
  3.80 & -0.476 & -0.778 & -1.552 & -2.562 \\
  3.83 & 9.74E-2 & 1.74E-1 & -0.562 & -0.998 \\
  \hline\hline
\end{tabular}
\end{table}

\bibliography{mybib}
\end{document}